\documentclass[conference]{IEEEtran}
\usepackage[T1]{fontenc}
\usepackage[utf8]{inputenc}
\usepackage{units}
\usepackage{amsmath}
\usepackage{amsthm}
\usepackage{amssymb}
\usepackage[bookmarks=true,bookmarksnumbered=true,bookmarksopen=true,bookmarksopenlevel=1,
 breaklinks=false,pdfborder={0 0 0},pdfborderstyle={},backref=false,colorlinks=false]
 {hyperref}
\hypersetup{pdftitle={Your Title},
 pdfauthor={Your Name},
 pdfpagelayout=OneColumn, pdfnewwindow=true, pdfstartview=XYZ, plainpages=false}

\makeatletter
\theoremstyle{plain}
\newtheorem{thm}{\protect\theoremname}
\theoremstyle{plain}
\newtheorem{prop}[thm]{\protect\propositionname}
\theoremstyle{definition}
\newtheorem{example}[thm]{\protect\examplename}
\theoremstyle{plain}
\newtheorem{cor}[thm]{\protect\corollaryname}
\theoremstyle{plain}
\newtheorem{lem}[thm]{\protect\lemmaname}
\theoremstyle{definition}
\newtheorem{defn}[thm]{\protect\definitionname}

\usepackage[caption=false,font=footnotesize]{subfig}

\makeatother

\providecommand{\corollaryname}{Corollary}
\providecommand{\definitionname}{Definition}
\providecommand{\examplename}{Example}
\providecommand{\lemmaname}{Lemma}
\providecommand{\propositionname}{Proposition}
\providecommand{\theoremname}{Theorem}

\begin{document}
\title{An Information Theoretic Proof of the Radon-Nikodym Theorem}
\author{\IEEEauthorblockN{Peter~Harremoës}\IEEEauthorblockA{Niels Brock\\
Copenhagen Business College\\
Copenhagen\\
Denmark\\
E-mail: harremoes@ieee.org}}
\maketitle
\begin{abstract}
The Radon-Nikodym theorem plays a significant role in the definition
of Shannon entropy, f-divergences, and other basic quantities in information
theory. The existence of Radon Nikodym derivates appear in many text
books in measure theory but in text books on probability or information
theory it is often omitted because the proof is often considered
to be too difficult. 
\end{abstract}

\IEEEpeerreviewmaketitle{}

\section{Introduction}

One of the fundamental tools in measure theory is the Radon-Nikodym
derivative, that allow us to describe measures as functions. For instance,
most continuous probability distributions are given in terms of their
density functions. In information theory, important concepts like differential
entropy and information divergence are usually defined using density
functions. Radon-Nikodym derivatives also play an important role in
various theoretical derivations. For instance, the existence of regular
conditional expectation is often based on the Radon-Nikodym Theorem.
Many expositions in information theory and probability theory do include
a proof of this important result. As we shall see in this short note,
one can prove the Radon-Nikodym using information projections. This
also led to an improved version of the Radon-Nikodym that not
only states the existence of Radon-Nikodym derivatives, but can also
quantify how close a finite approximation is to the Radon-Nikodym derivative.

In information theory, uncertainty is usually quantified using probability
theory as it was developed by Kolmogorov. For instance, a simple information
source is given in terms of an alphabet with a set of non-negative weights
called probabilities that add up to 1. In many cases, it will simplify
the computations if the constraint on the total mass is dropped. In
general, we get a more flexible language for modelling uncertainty
if we allow them to use measures that are not normalized.

Recently, a theory of expectation measures was introduced as an alternative
to the usual Kolmogorov style of probability theory \cite{Harremoes2025a,Harremoes2025b}.
This new approach allows us to distinguish between different applications
of measure theory when it is used to model randomness or uncertainty.
Some of these measures have total mass 1, and some have finite total
mass greater than 1 or less than 1. Sometimes the measures may even
have infinite total mass. 

To some extend our exposition even works if the measures $\mu$
and $\nu$ are replaced by \emph{valuations}, i.e. set functions defined
on a distributive lattice rather than on a $\sigma$-algebra. The
relevance of working with valuations is discussed in \cite[Sec. 3.1-3.3]{Harremoes2025a}.

\section{Lattices and valuations}

Usually, results in information theory and statistics are formulated
in terms of probability theory as it was formulated by Kolmogorov.
In \cite{Harremoes2025a}, it was demonstrated how a theory of uncertainty
can be based on expectation measures. These measures differ from probability
measures in that there is no requirement that the measures are normalized
so that the total mass is 1. Expectation measures can be viewed as
measures that quantify the expected number of observations in a point
process. In this paper, we will go one step further and replace measures
on $\sigma$-algebras by valuations on lattices. This requires some
motivation that will be given in the present section.

\subsection{Shannon inequalities on lattices of functional dependencies}

Let $X$ denote a random variable. Then the entropy of $X$ is given
by
\[
H\left(X\right)=-\sum_{x}P\left(X=x\right)\ln\left(P\left(X=x\right)\right).
\]
If $S=\left\{ X_{1},X_{2},\dots,X_{n}\right\} $ then 
\[H\left(S\right)=H\left(\left(X_{1},X_{2},\dots,X_{n}\right)\right).
\]
For a set of variables, the entropy function is a function from the
powerset of the set of variables to the real numbers. The power set
is a lattice with intersection $\cap$ and union $\cup as$lattice
operations. The entropy function satisfies the three Shannon inequalities.

\textbf{Strictness} $H\left(\emptyset\right)=0.$

\textbf{Monotonicity} If $S\subseteq T$ then $H\left(S\right)\leq H\left(T\right).$

\textbf{Submodularity} $H\left(S\right)+H\left(T\right)\geq H\left(S\cap T\right)+H\left(S\cup T\right).$

In addition to these inequalities, the entropy function also satisfies
so-called non-Shannon inequalities \cite{Zhang1997}, which are inqualities
that cannot be derived directly from the Shannon inequalities. As
noted by R. Yeung \cite{Yeung2002} the entropy function behaves much
like a measure, and this is a direct consequence of the three Shannon
inequalities. 

If $S\subseteq T$ and $H\left(S\right)=H\left(T\right)$ then each
variable in $T$ is a function of the variables in $S$ and we say
that the variables in $T$ are determined by the variables in $S$.
Let $S$ be a set of variables. Then we introduce the functional dependency
closure of $S$ by as the set 
\[
cl\left(S\right)=\bigcup T
\]
where the union is taken over all $T$ determined by $S$. Then the
set of closed sets of variables forms a lattice with $\cap$ as meet
operator and with join operator $\uplus$ defined by $S\uplus T=cl\left(S\cup T\right).$
This lattice is called the functional dependence lattice. The restriction
of the entropy function to the closed sets of variables is again a
function that satisfies positivity, monotonicity and submodularity,
where $\cup$ operator in the submodular inequality is replaced by
$\uplus$ operator. Entropy on lattices has been studied in more
detail in \cite{Harremoes2015,Harremoes2018a}.

Entropy inequalities are usually discussed in terms of sets of variables,
but for the inequalities, we are not really interested in the values
of random variables. Instead, we are interested in the $\sigma$-algebras
generated by the variables. If $S_{i}$ generated the $\sigma$-algebra
$\mathcal{F}_{i}$ then $S_{1}\cap S_{2}$ generates the $\sigma$-algebra
$\mathcal{F}_{1}\cap\mathcal{F}_{2}$, and $S_{1}\uplus S_{2}$ generates
the $\sigma$-algebra $\sigma\left(\mathcal{F}_{1}\cap\mathcal{F}_{2}\right),$
i.e. the smallest $\sigma$-algebra generated by $\mathcal{F}_{1}$
and $\mathcal{F}_{2}$. Thus, the Shannon inequalities can be formulated
in terms of a $\sigma$-algebra and a system of subalgebras that form
a lattice.

\subsection{Concept lattices}

We consider a situation where some objects $g\in G$ are classified
according to their properties. For each object $g\in G$ and each
property $m\in M$ we write $gIm$ if object $g$ has property $m.$
Let $X\subseteq G$ and $Y\subseteq M.$ Then define
\begin{align*}
X' & =\left\{ m\in M\mid gIm\textrm{ for all }g\in X\right\} ,\\
Y' & =\left\{ g\in G\mid gIm\textrm{ for all }m\in Y\right\} .
\end{align*}
A formal concept is defined as a pair $\left(X,Y\right)$ such that
$X=Y'$ and $Y=X'.$ If $\left(X,Y\right)$ is a formal concept then
$X''=X$ and $Y''=Y.$ The mapping $X\to X''$ is a closure operator
and the formal concepts may be identified with closed sets of objects
under this closure operator. The closed sets of objects are ordered
by inclusion. Under this ordering the closed sets form a complete lattice
with the following lattice operations.
\begin{align*}
\bigwedge_{i}X_{i} & =\bigcap_{i}X_{i},\\
\bigvee_{i}X_{i} & =\left(\bigcup X_{i}\right)''.
\end{align*}

Let $\left(L,\leq\right)$ be a complete lattice. Let $G=L$ and let
$M=L.$ Let $\ell_{1},\ell_{2}\in L.$ If $\ell_{1}I\ell_{2}$ if
and only if $\ell_{1}\leq\ell_{2}$ then $L$ is equivalent to the
concept lattice generated by $I.$ In this way any complete lattice
can be represented as a concept lattice. In particular any finite
lattice has the structure as a concept lattice. 

Let $G$ be a set of objects and let $\mu\left(X\right)$ denote the number of elements
in $X.$ Then $\mu$ is \emph{modular}, i.e.
\begin{equation}
\mu\left(X_{1}\right)+\mu\left(X_{2}\right)=\mu\left(X_{1}\cap X_{2}\right)+\mu\left(X_{1}\cup X_{2}\right).    
\end{equation}
If $M$ is a set of properties then for $X_1 , X_2 \subseteq G$ we have 
$\mu\left(X_1\cup X_2\right)\leq\mu\left(X_1\vee X_2\right)$. 
Therefore, for closed subsets of $G$ we have:

\textbf{Strictness} $\mu\left(\emptyset\right)=0$.

\textbf{Monotonicity} If $X_{1}\leq X_{2}$ then $\mu\left(X_{1}\right)\leq\mu\left(X_{2}\right).$

\textbf{Super-modularity} $\mu\left(X_{1}\right)+\mu\left(X_{2}\right)\leq\mu\left(X_{1}\wedge X_{2}\right)+\mu\left(X_{1}\vee X_{2}\right).$

Let $\mathcal{L}$ denote a finite lattice and let $\mu$ denote a function that is strict, 
monotone and super-modular. If $x,a,b\in \mathcal{L}$ and $a\leq x$ and $b\leq x$ and 
$\mu(a)=\mu(b)=\mu(x)$ then the super-modular inequality implies that 
$\mu(x)=\mu(a\wedge b)$. Therefore there exists a smallest element $y$ in the lattice such 
that $y\leq x$ and $\mu(y)=\mu(x)$. This element will be denoted $cocl(x)$. The operator 
$cocl$ is a co-closure operator in the lattice, i.e. it is a closure in the lattice equipped 
with the reverse ordering. For a concept lattice with $A\subseteq M$ we have that $cocl(A')=A''$. In any lattice the subset of co-closed elements in the lattice form a new lattice 
with the same ordering as the original lattice. 
The restriction of the super-modular function $\mu$ to the lattice of co-closed elements is 
again a super-modular function. By construction the restricted function is strictly monotone.

\subsection{Valuations and distributive lattices}

Let $\left(L,\wedge,\vee\right)$ denote a lattice. Then a \emph{valuation}
on $\left(L,\wedge,\vee\right)$ is defined as a function $\mu:L\to\left[0,\infty\right]$ that
satisfies strictness, monononicity, and modularity.
\begin{prop}
Let $\left(L,\wedge,\vee\right)$ denote a lattice with a valuation $\mu$. The set 
of co-closed elements is a modular lattice.
\end{prop}
\begin{IEEEproof}
    The restriction of $mu$ to the lattice of co-closed elements is strictly monotone. 
    Therefore we may assume that $\mu$ is strictly monotone.
    
    Assume that $X\leq Z $. We have to demonstrate that 
    $\left(X\vee Y\right)\wedge Z \supseteq X\vee\left(Y\wedge Z\right)$ 
    holds with equality. We have 
    \begin{equation}
    \begin{split}
        \mu\left(X\vee\left(Y\wedge Z\right)\right)
        & = \mu\left(X\right)+\mu\left(Y\wedge Z\right)-\mu\left(X\wedge Y\wedge Z\right)\\
        & =\mu\left(X\right)+\mu\left(Y\wedge Z\right)-\mu\left(X\wedge Y\right).
    \end{split}
    \end{equation}
    We also have 
    \begin{equation}
        \begin{split}
            \mu\left(\left(X\vee Y\right)\wedge Z\right) 
            = &\mu\left(X\vee Y\right) +\mu\left(Z\right)-\mu\left(X\vee Y\vee Z\right)\\
             =& \mu\left(X\vee Y\right) +\mu\left(Z\right)-\mu\left(Y\vee Z\right)\\
             =& \mu\left(X\right)+\mu\left(Y\right)-\mu\left(X\wedge Y\right)+\mu\left(Z\right)\\
            &-\left(\mu(Y)+\mu(Z)-\mu\left(Y\wedge Z\right)\right)\\
            =&\mu\left(X\right)+\mu\left(Y\wedge Z\right)-\mu\left(X\wedge Y\right).
        \end{split}
    \end{equation}
    The equation 
    \begin{equation}
        \mu\left(\left(X\vee Y\right)\wedge Z\right)
        = \mu\left(X\vee\left(Y\wedge Z\right)\right)
    \end{equation}
    together with strict monotonicity implies that $\left(X\vee Y\right)\wedge Z
        = X\vee\left(Y\wedge Z\right).$
\end{IEEEproof}

\begin{prop}
Let $G$ denote a set of objects and let $M$ denote a set of properties.
If $\mu$ denotes the number of elements in a set of objects 
and $\mu$ is a valuation, then the lattice is distributive.
\end{prop}
\begin{IEEEproof}
We have 
\[
\mu\left(X_{1}\right)+\mu\left(X_{2}\right)=\mu\left(X_{1}\cap X_{2}\right)+\mu\left(X_{1}\cup X_{2}\right)
\]
and $X_{1}\wedge X_{2}=X_{1}\cap X_{2}.$ Hence, $\mu\left(X_{1}\cup X_{2}\right)=\mu\left(X_{1}\vee X_{2}\right).$
Since, $X_{1}\vee X_{2}$ is the closure of $X_{1}\cup X_{2}$ we
must have $X_{1}\vee X_{2}=X_{1}\cup X_{2}.$ Therefore the concept 
lattice is a sub-lattice of the power set lattice, which is
distributive.
\end{IEEEproof}
If $j$  is an irreducible element in a lattice then we will use $j^-$ 
to denote the unique element that covers $j$.
\begin{prop}\label{prop:kaede}
    Let $\left(L,\wedge ,\vee\right)$ be a distributive lattice of rank $r$. Then 
    for any maximal chain $m_0 < m_1 <\dots < m_r$ there exists a sequence of 
    irreducible elements $j_1 ,j_2 , \dots ,j_r$  such that $m_{i+1}=m_i \vee j_i$ 
    and $m_i\wedge j_i= j_i ^-$.
\end{prop}
\begin{IEEEproof}
    The lattice can be represented as the downsets of the ordered set of irreducible 
    elements. In particular, an irreducible element can be represented as a maximal chain 
    from the empty set to the irreducible element. Then, a maximal chain in the lattice is 
    obtained by starting by the empty set and adding one irreducible element at a time. 
\end{IEEEproof}

In the sequel, we shall focus on valuations defined on distributive
lattices. The lattices work as classification systems. The valuation
could be used to quantify observed data points or to quantify the
expected number data points. For observations the valuations will
have values in $\mathbb{N}_{0}\cup\left\{ \infty\right\} $. 

Let $\left(L,\wedge,\vee\right)$ denote a distributive lattice. A
lattice element $j\in L$ is said to be $\vee$-irreducible if $j=a\vee b$
implies $a\leq b$ or $b\leq a.$ Let $J$ denote the set of $\vee$-irreducible
elements. The set $J$ is ordered by the same ordering as the lattice.
If $g\in J$ and $m\in L$ then a relation between $J$ and $L$ is
given by $g\leq m.$ For any element $m\in L$ we get a downset $m'=\left\{ g\in J\mid g\leq m\right\} ,$
and since the lattice is distributive, all downsets have this form.
Thus, any finite distributive lattice can be represented as the downsets
of a poset. The set of downsets forms a sublattice of the powerset
of the set $J$. According to the Birkhoff-Stone Theorem, distributive
lattices have nice representations as sub-lattices of a Boolean lattice.
In the subsequent sections, we will only use this result for finite
distributive lattices where the theory is quite simple.
\begin{example}
Let $\left(\mathbb{A},\tau\right)$ denote a topological space with
$\tau$ denotes the set of open sets. Then $\tau$ is a distributive
lattice. The lattice $\tau$ is not Boolean because the complement
of an open set is, in general, not open. The lattice $\tau$ can be
embedded in the Borel $\sigma$-algebra, which is a Boolean algebra.
\end{example}
A topology is a frame, i.e. it is a lattice where arbitrary joins
are allowed and where meet is distributive over joins. In the present
paper, we will focus on valuations that are continuous in the sense
that for any directed net $x_{\lambda}$ in the lattice, we have

\[
\mu\left(\bigvee_{\lambda}x_{\lambda}\right)=\sup_{\lambda}\mu\left(x_{\lambda}\right).
\]
This property replaces $\sigma$-additivity and inner regularity in
measure theory.

\begin{example}
Any finite continuous valuation on a compact Hausdorff space extends
uniquely to a regular $\tau$-smooth Borel measure.
\end{example}
\begin{example}
Any locally finite continuous valuation on a metric space extends
uniquely to a regular $\tau$-smooth Borel measure.
\end{example}
For more general topological spaces it is an open question to what
extend all valuations are given by measures. In the rest of this paper
we will formulate our results for topological spaces although our
setting allow more general versions of our theorems.

If $A\in\tau$ is an open set then $1_{A}$ is lower semi-continuous
function. Any function of the form $f=\sum_{i}c_{i}1_{A_{i}}$ is
also lower semi-continuous as long as the coefficients are positive
numbers. Using modularity of a valuation one can prove that one can
define an integral with respect to an evaluation $\mu$ by
\[
\int f\,\mathrm{d}\text{\ensuremath{\mu}}=\sum_{i}c_{i}\mu\left(A_{i}\right).
\]
Any lower semi-continuous function $g$ can be approximated from below
by linear combinations of indicator functions so we can define
\[
\int g\,\mathrm{d}\mu=\sup_{f\leq g}\int f\,\mathrm{d}\mu.
\]
The completion of the linear span of lower semi-continuous function
is the set of $L^{1}\left(\mathbb{A},\mu\right).$ The last part of
the construction is pretty standard in functional analysis \cite[Sec. 6.3]{Pedersen2012}.

\section{Information divergences\label{sec:Information-divergences}}

Let $\mathcal{K}$ be a finite distributive lattice with valuations $\mu$ and $\nu$. If $j$ 
is an irreducible element in $\mathcal{K}$ then $j^-$ will be used to denote the element 
in the lattice that is covered by $j$. Let $\Delta \mu$ denote the function 
$\Delta \mu (j )=\mu(j)-\mu(j^-)$ for $i=2,3,\dots,n$. Then the information divergence can be defined by
\[
D\left(\mu\Vert\nu\right)=\begin{cases}
\sum_{j} f\left(\frac{\Delta\mu\left(j\right)}{\Delta\nu\left(j\right)}\right)\Delta\nu\left(j\right) & \textrm{if }\mu\preceq\nu\\
\infty & \textrm{else}
\end{cases}
\]
where the sum is taken over all irreducible elements and $f$ is the function $f\left(x\right)=x\ln\left(x\right)-\left(x-1\right)$
and we use the conventions that $f\left(\frac{\infty}{\infty}\right)\cdot\infty=0$ 
and $f\left(\frac{0}{0}\right)\cdot 0=0$. The divergence $D\left(\mu\Vert\nu\right)$ 
is equal the divergence of $\mu$ and $\nu$ extended to the $\sigma$-algebra generated by the lattice $\mathcal{K}$.
\begin{prop}
    Let $\mathcal{L}$ be a finite distributive lattice and let $\mathcal{K}$ be a maximal chain in $\mathcal{L}$. If $\mu$ and $\nu$ are valuations on $\mathcal{L}$ then
    \begin{equation}
        D\left(\mu\Vert\nu\right)=D\left(\mu_{\mid\mathcal{K}}\Vert\nu_{\mid\mathcal{K}}\right)
    \end{equation}
\end{prop}
\begin{IEEEproof}
        Let the chain $\mathcal{K}$ be given by a sequence $m_0 <m_1 <\dots <m_r$. 
        Then $m_i$ is an irreducible element in $\mathcal{K}$ for $i=1,2,\dots ,r$. 
        According the Proposition \ref{prop:kaede} there exists a sequence of 
        irreducible elements $j_1 ,j_2 ,\dots ,j_r$ such that 
        $m_i \vee j_i =m_{i+1}$ and $m_i\wedge j_i =j_i^-$. The modular equation implies that 
        \begin{equation}
            \mu\left(m_{i+1}\right)-\mu\left(m_i\right)=\mu\left(j_i\right)-\mu\left(j_i^-\right).
        \end{equation}
\end{IEEEproof}
\begin{cor}
    Let $\mu$ and $\nu$ denote valuations on a finite distributive lattice. Then the the divergence of $\mu$ from $\nu$ equals the divergence of $\mu$ from $\nu$ extended to measures on the $\sigma$-algebra generated by tha lattice. 
\end{cor}
From the previous proposition it follows that in a distributive lattice all maximal chains have the same diveregence. Our next proposition states that this holds for any lattice.   
\begin{prop}
    Let $\left(L,\wedge,\vee\right)$ denote a finite lattice with valuations $\mu$ and $\nu$. Then all maximal chains have the same information divergence.
\end{prop}
\begin{IEEEproof}
    Recall that the rank of a lattice is the length of the longest chain in the lattice. The proof goes by induction in the rank $r$ of the lattice.
    
    For a lattice of rank $r=1$ the result is trivial.
    
    Assume that the result holds for all lattices of rank up to $r=\ell$ and 
    assume that $\left(L,\wedge,\vee\right)$ has rank $r=\ell+1$. Consider two 
    maximal chains $a_1 ,a_2 ,\dots a_m ,a_{m+1}$ and $b_1 ,b_2 ,\dots b_n ,b_{n+1}$ 
    in $\left(L,\wedge,\vee\right)$. Note that $a_{m+1} =b_{n+1}$. Let $c=a_m \wedge b_n$. 
    Then the divergence of the chain $a_1 ,a_2 ,\dots a_m ,a_{m+1}$ is identical 
    to the divergence of a chain through both $c$ and $a_m$. Similarly, the 
    divergence of the chain $b_1 ,b_2 ,\dots b_n ,b_n+1$ is identical to the 
    divergence of a chain through both $c$ and $b_n$. We may assume that these 
    two chains are identical below $c$, so that $c=a_k =b_k$. If $a_m =b_n$ we 
    have $c=a_m =b_n$ and we are done. Assume that $a_m \neq b_n$. In this case 
    $a_m \vee b_n =a_{m+1}$, and modularity leads to
    \begin{equation*}
        \begin{split}
            \mu\left(a_{m+1}\right)+\mu(c)&=\mu\left(a_m\right)+\mu\left(b_n\right),\\
            \mu\left(a_{m+1}\right)-\mu\left(a_m\right)&=\mu\left(b_n\right)-\mu(c),\\
            \mu\left(a_{m+1}\right)-\mu\left(b_n\right)&=\mu\left(a_m\right)-\mu(c).
        \end{split}
    \end{equation*}
    By the construction we also have $a_{k+1}\wedge b_{k+1}=c$ and $a_{k+1}\vee b_{k+1}=a_{n+1}$. Hence
    \begin{equation}
        \begin{split}
            \mu\left(a_{m+1}\right)-\mu\left(a_{m}\right)&=\mu\left(b_{k+1}\right)-\mu\left(b_k\right),\\
            \mu\left(b_{m+1}\right)-\mu\left(b_{m}\right)&=\mu\left(a_{k+1}\right)-\mu\left(a_k\right).
        \end{split}
    \end{equation}
    In particular, $\mu\left(a_{k+1}\right)=\mu\left(a_m\right)$  
    and $\mu\left(b_{k+1}\right)=\mu\left(b_n\right)$. Thus, 
    $\Delta\mu\left(a_{m+1}\right)=\Delta\mu\left(b_{k+1}\right)$ and 
    $\Delta\mu\left(b_{n+1}\right)=\Delta\mu\left(a_{k+1}\right)$ and all other 
    increments in the two chains between $c$ and the maximal element are zero.
\end{IEEEproof}

If
$\mu$ and $\nu$ are valuations on a general distributive lattice,
then the information divergence is defined by 
\[
D\left(\mu\Vert\nu\right)=\sup_{\mathcal{K}}D\left(\mu_{\mid\mathcal{K}}\left\Vert \nu_{\mid K}\right.\right),
\]
where the supremum is taken over all finite sublattices $\mathcal{K}\subseteq\mathcal{L}$
for which $\emptyset\in\mathcal{K}.$  

An important property of information divergence is \emph{homogenuity},
i.e. for any $t\geq0$ we have $D\left(t\cdot\mu\Vert t\cdot\nu\right)=t\cdot D\left(\mu\Vert\nu\right).$
According to \emph{Gibbs' inequality} $D\left(\mu\Vert\nu\right)\geq0$
holds with equality if and only if $\mu=\nu.$

A valuation $\mu$ on the distributive lattice $\mathcal{L}$ is said to be
$\sigma$-finite if there exists a sequence $A_{n}\in \mathcal{L}$ such that
$\mu\left(A_{n}\right)<\infty$ for all $n$ and for all $B\in \mathcal{L}$
we have $\mu\left(A_{n}\wedge B\right)\to\text{\ensuremath{\mu}\ensuremath{\left(B\right)} for}$
$n\to\infty.$ 
\begin{prop}
\label{prop:Decomp}Let $\mu$ and $\nu$ denote valuations on the
distributive lattice $\mathcal{L},$ and assume that $D\left(\mu\Vert\nu\right)<\infty.$
Then there exists a sublattice $\mathcal{K}\subseteq\mathcal{L}$
such that $\mu_{\mid K}$ and $\nu_{\mid K}$ are $\sigma$-finite
and such that $D\left(\mu\Vert\nu\right)=D\left(\mu_{\mid\mathcal{K}}\Vert\nu_{\mid\mathcal{K}}\right)$. 
\end{prop}
\begin{IEEEproof}
Let $\mathcal{K}_{1}\subseteq\mathcal{K}_{2}\subseteq\dots\subseteq\mathcal{L}$
denote a sequence of sub-lattices such that $D\left(\mu_{\mid\mathcal{K}_{j}}\left\Vert \nu_{\mid\mathcal{K}_{j}}\right.\right)\to D\left(\mu\left\Vert \nu\right.\right)$
for $j\to\infty.$ Let $\mathcal{K}=\bigcup_{n}\mathcal{K}_{n}$.
Each sublattice $K_{n}$ has a maximal element $A_{n}$ for which
$\text{\ensuremath{\mu}\ensuremath{\left(A_{n}\right)}<\ensuremath{\infty}. Since }$$D\left(\mu_{\mid\mathcal{K}_{j}}\left\Vert \nu_{\mid\mathcal{K}_{j}}\right.\right)<\infty$
we must have that $\mu\left(B\right)=\infty\Leftrightarrow\nu\left(B\right)=\infty.$
Let $\mathcal{M}_{n}=\left\{ B\in\mathcal{K}_{n}\mid B\leq A_{n}\right\} .$
Then $D\left(\mu_{\mid\mathcal{M}_{j}}\left\Vert \nu_{\mid\mathcal{M}_{j}}\right.\right)=D\left(\mu_{\mid\mathcal{K}_{j}}\left\Vert \nu_{\mid\mathcal{K}_{j}}\right.\right).$
Let $\mathcal{K}=\bigcup_{n}\mathcal{M}_{n}$. Then $\mu_{\mid K}$
and $\nu_{\mid K}$ are $\sigma$-finite and $D\left(\mu\Vert\nu\right)=D\left(\mu_{\mid\mathcal{K}}\Vert\nu_{\mid\mathcal{K}}\right).$
\end{IEEEproof}
\begin{cor}
\label{cor:sfinite}Assume that $\mu$ and $\nu$ are measures such
that $D\left(\mu\Vert\nu\right)<\infty.$ If $\mu$ is s-finite then
$\nu$ is s-finite.
\end{cor}
\begin{IEEEproof}
There exists a measurable set $B$ such that both $\mu\left(\cdot\cap B\right)$
and $\nu\left(\cdot\cap B\right)$ are $\sigma$-finite and such that
$\mu\left(\cdot\cap\complement B\right)=\nu\left(\cdot\cap\complement B\right).$
Therefore $\nu\left(\cdot\cap B\right)$ is $\sigma$-finite and $\nu\left(\cdot\cap\complement B\right)=\mu\left(\cdot\cap\complement B\right).$
Since $\nu$ is s-finite, we see that $\mu\left(\cdot\cap\complement B\right)$
is s-finite and since $\mu\left(\cdot\cap B\right)$ is $\sigma$-finite
we conclude that $\mu$ is s-finite.

The above exposition is more general than what is presented in the
recent paper \cite{Leskelae2024}, which is limited to $\sigma$-finite
measures.
\end{IEEEproof}
Let $\mu$ and $\nu$ denote valuations on the distributive lattice
$\mathcal{L},$ and assume that $D\left(\mu\Vert\nu\right)<\infty.$
Let $M_{n}$ be a finite sublattice of $L$ such that $\mu$ and $\nu$
are finte on $M_{n}.$ $M_{n}$ generates a finite $\sigma$-algebra
and one can define the measurable function $f_{n}$ by
\[
f_{n}\left(a\right)=\begin{cases}
\frac{\mu\left(a\right)}{\nu\left(a\right)}, & \textrm{if }a\textrm{ is an atom of }M_{n};\\
1, & \textrm{else.}
\end{cases}
\]
 Then a measure $\mu_{n}$ can be defined by 
\[
\mu_{n}\left(A\right)=\int_{A}f_{n}\left(a\right)\,\mathrm{d}\nu.
\]
It is easy to check that $D\left(\mu_{n}\Vert\nu\right)=D\left(\mu_{\mid M_{n}}\Vert\nu_{\mid M_{n}}\right).$ Further
we have that 
\[
D\left(\theta\Vert\nu\right)=D\left(\theta\Vert\mu_{n}\right)+D\left(\mu_{n}\Vert\nu\right)
\]
 So that 
\[
D\left(\mu_{n}\Vert\nu\right)\leq D\left(\theta\Vert\nu\right)
\]
 for all valuations $\theta$ such $\theta_{\mid M_{n}}=\mu_{\mid M_{n}}.$ Therefore
$\mu_{n}$ can be considered as an \emph{information projection} of
$\nu$ on the set of valuations $\left\{ \theta\mid\theta_{\mid M_{n}}=\mu_{\mid M_{n}}\right\} .$
\begin{prop}
Let $\mu$ and $\nu$ denote valuations on the distributive lattice
$\mathcal{L},$ and assume that $D\left(\mu\Vert\nu\right)<\infty,$
and let $M_{n}$ be sequence of finite sublattice of $L$ such that
$\mu$ and $\nu$ are finte on $M_{n}.$ and such that $D\left(\mu_{\mid M_{n}}\Vert\nu_{\mid M_{n}}\right)\to D\left(\mu\Vert\nu\right)$
for $n\to\infty.$ Then $\mu_{n}$ converges to $\mu$ setwise, i.e.
for any $A\in L$ we have $\mu_{n}\left(A\right)\to\mu\left(A\right).$
\end{prop}
\begin{IEEEproof}
Since $D\left(\mu_{\mid M_{n}}\Vert\nu_{\mid M_{n}}\right)\to D\left(\mu\Vert\nu\right)$
we have $D\left(\mu\Vert\mu_{n}\right)\to0$ for $n\to\infty.$ By
the data processing inequality, we have
\[
D\left(\mu\left(A\right)\Vert\mu_{n}\left(A\right)\right)\to0
\]
If $\mu\left(A\right)<\infty$ this implies that $\mu_{n}\left(A\right)\to\mu\left(A\right).$
If $\mu\left(A\right)=\infty$ it implies that $\mu_{n}\left(A\right)=\infty$
eventually.
\end{IEEEproof}
Next we will demostrate that the sequence of functions $f_{n}$ converges
to a measurable function, but since the measures are not not normalized
some care is needed in how this result is formulated.
\begin{lem}
Let $\mu$ and $\nu$ denote finite valuations on the topological space
$\left(\mathbb{A},\tau\right)$ with $D\left(\mu\Vert\nu\right)<\infty$
Then $f_{n}$ is a Cauchy sequence in $L_{1}\left(\mathbb{A},\nu\right),$
and $f_{n}$ convergences to a function $\rho$ such that $\int f\,\mathrm{d}\mu=\int f\rho\,\mathrm{d}\nu.$
\end{lem}
\begin{IEEEproof}
For $m\leq n$ we have 
\begin{align*}
D\left(\mu_{n}\Vert\mu_{m}\right) & =D\left(\mu_{n}\Vert\nu\right)-D\left(\mu_{m}\Vert\nu\right)\\
 & \leq D\left(\mu\Vert\nu\right)-D\left(\mu_{m}\Vert\nu\right).
\end{align*}
Now
\begin{align*}
D\left(\mu_{n}\Vert\mu_{m}\right) & =\left|\mu\right|\cdot D\left(\frac{\mu_{n}}{\left|\mu\right|}\left\Vert \frac{\mu_{m}}{\left|\mu\right|}\right.\right)\\
 & \geq\left|\mu\right|\cdot\frac{\left|\frac{\mu_{n}}{\left|\mu\right|}-\frac{\mu_{m}}{\left|\mu\right|}\right|^{2}}{2}\\
 & =\frac{\left|\mu_{n}-\mu_{m}\right|^{2}}{2\left|\mu\right|}\\
 & =\frac{\left\vert f_{n}-f_{m}\right\vert ^{2}}{2\left|\mu\right|}.
\end{align*}
Therefore $\left\vert f_{n}-f_{m}\right\vert \to0$ for $m,n\to\infty,$
and since $f_{n}$ is a Cauchy sequence there exists a measurable function
$\rho$ such that $\left\Vert f_{n}-\rho\right\Vert \to0$ for $m,n\to\infty,$
Since $\int_{A}f_{n}\,\mathrm{d}\nu=\mu_{n}\left(A\right)\to\mu\left(A\right)$
and $\int_{A}f_{n}\,d\nu\to\int_{A}\rho\,\mathrm{d}\nu$ we have 
\[
\mu\left(A\right)=\int_{A}\rho\,\mathrm{d}\nu.
\]
\end{IEEEproof}
\begin{lem}
Let $\mu$ and $\nu$ denote $\sigma$-finite measures on the topological
space $\left(\mathbb{A},\tau\right)$ with $D\left(\mu\Vert\nu\right)<\infty.$
Then $f_{n}$ is a Cauchy sequence with respect to the norm
\[
\text{\ensuremath{\left\Vert f-g\right\Vert }=\ensuremath{\sum_{i=1}^{\infty}\frac{1}{2^{i}}\frac{\left\Vert f-g\right\Vert _{\nu_{\mid\mathcal{M}_{i}}}}{1+\left\Vert f-g\right\Vert _{\nu_{\mid\mathcal{M}_{i}}}}},}
\]
and $f_{n}$ convergences to a function $\rho$ such that $\int f\,\mathrm{d}\mu=\int f\rho\,\mathrm{d}\nu.$
\end{lem}

\section{Pointwise convergence}

Most the the results in Section \ref{sec:Information-divergences}
can be carried out with other $f$-divergences than information divergence.
We have seen that the restricted densities converge to the Radon-Nikodym
derivative in $L^{1}$ sense. This implies that there exists a subsequence
that converges almost surely. In Subsection 5.1 we will prove that
a bound on information divergence implies convergence of the whole
sequence almost surely, and in Subsection 5.2 we will prove that a
bound on information divergence is necessary.

\subsection{Almost sure pointwise convergence}
\begin{lem}[{Doob's Maximal Inequalitites \cite[p. 494]{Shiryaev1996}}]
Let $\left(X_{1},F_{1}\right),\left(X_{2},F_{2}\right),\dots,\left(X_{n},F_{n}\right)$
denote a non-negative martingale with respect to the probability measure
$Q$. Let $X^{max}=\max_{j=1,\dots,n}X_{j}$ and $X^{min}=\min_{j=1,\dots,n}X_{j}$.
Then
\begin{align*}
\lambda\cdot Q\left(X^{max}\geq\lambda\right) & \leq E\left(X_{n}\cdot1_{X^{max}\geq\lambda}\right),\\
\lambda\cdot Q\left(X^{min}\geq\lambda\right) & \geq E\left(X_{n}\cdot1_{X^{min}\geq\lambda}\right).
\end{align*}
\end{lem}
\begin{thm}[\cite{Harremoes2008b}]
 Let $\left(X_{1},F_{1}\right),\left(X_{2},F_{2}\right),\dots,\left(X_{n},F_{n}\right)$
denote a non-negative martingale with respect to the probability measure
$Q$. Let $X_{m,n}^{max}=\max_{j=m,\dots,n}X_{j}$ and $X_{m,n}^{min}=\min_{j=m,\dots,n}X_{j}$.
Assume that $E\left[X_{n}\right]=1.$ Then 
\begin{align*}
\gamma\left(E\left[X_{m,n}^{max}\right]\right) & \leq D\left(\left.P_{n}\right\Vert P_{m}\right),\\
\gamma\left(E\left[X_{m,n}^{min}\right]\right) & \leq D\left(\left.P_{n}\right\Vert P_{m}\right),
\end{align*}
where $P_{j}$ is the probability measure given by $\frac{\mathrm{d}P_{j}}{\mathrm{d}Q}=X_{j},$
and $\gamma\left(x\right)=x-1-\ln\left(x\right).$
\end{thm}
\begin{lem}
\label{lem:MinMax}Let $\mu$ and $\nu$ denote evaluations such that
$D\left(\mu\Vert\nu\right)<\infty$ and that $k=\left\vert \mu\right\vert <\infty.$
Let $F_{1}\subseteq F_{2}\subseteq\dots\subseteq F_{n}$. Let $Y_{j}=\frac{\mathrm{d}\mu\left(\cdot\mid F_{j}\right)}{\mathrm{d}\nu\left(\cdot\mid F_{j}\right)}$
and let where $\mu_{j}$ be the measure given by $\frac{\mathrm{d}\mu_{j}}{\mathrm{d}\nu}=X_{j}.$
and let $Y_{m,n}^{min}=\min_{m\leq j\leq n}Y_{j}$ and $Y_{m,n}^{max}=\max_{m\leq j\leq n}Y_{j}.$
Then 
\begin{align*}
D\left(k\left\Vert \int Y_{m,n}^{max}\,\mathrm{d}\nu\right.\right) & \leq D\left(\left.\mu_{n}\right\Vert \mu_{m}\right),\\
D\left(k\left\Vert \int Y_{m,n}^{min}\,\mathrm{d}\nu\right.\right) & \leq D\left(\left.\mu_{n}\right\Vert \mu_{m}\right).
\end{align*}
\end{lem}
\begin{IEEEproof}
Let $\ell=\left\vert \nu\right\vert .$ Then $P_{j}=\frac{\mu_{j}}{k}$
is a sequence of probability measures, and $Q=\frac{\nu}{\ell}$ is a probability
measure. Further $Y_{j}=\frac{\mathrm{d}\mu_{j}}{\mathrm{d}\nu}=\frac{k}{\ell}\cdot\frac{\mathrm{d}P_{j}}{\mathrm{d}Q}=\frac{k}{\ell}\cdot X_{j}$
so that $X_{j}=\frac{\ell}{k}Y_{j}.$ Then
\begin{align*}
\gamma\left(\int\frac{\ell}{k}Y_{m,n}^{max}\,\frac{\mathrm{d}\nu}{\ell}\right) & \leq D\left(\left.\frac{\mu_{n}}{k}\right\Vert \frac{\mu_{m}}{k}\right),\\
\gamma\left(\frac{1}{k}\int Y_{m,n}^{max}\,\mathrm{d}\nu\right) & \leq\frac{1}{k}D\left(\left.\mu_{n}\right\Vert \mu_{m}\right),\\
\frac{1}{k}\int Y_{m,n}^{max}\,\mathrm{d}\nu-1-\ln\left(\frac{1}{k}\int Y_{m,n}^{max}\,\mathrm{d}\nu\right) & \leq\frac{1}{k}D\left(\left.\mu_{n}\right\Vert \mu_{m}\right),\\
\int Y_{m,n}^{max}\,\mathrm{d}\nu-k-k\ln\left(\frac{1}{k}\int Y_{m,n}^{max}\,\mathrm{d}\nu\right) & \leq D\left(\left.\mu_{n}\right\Vert \mu_{m}\right),\\
k\ln\left(\frac{k}{\int Y_{m,n}^{max}\,\mathrm{d}\nu}\right)-\left(k-\int Y_{m,n}^{max}\,\mathrm{d}\nu\right) & \leq D\left(\left.\mu_{n}\right\Vert \mu_{m}\right),\\
D\left(k\left\Vert \int Y_{m,n}^{max}\,\mathrm{d}\nu\right.\right) & \leq D\left(\left.\mu_{n}\right\Vert \mu_{m}\right).
\end{align*}
The inequality $D\left(k\left\Vert \int Y_{m,n}^{min}\,\mathrm{d}\nu\right.\right)\leq D\left(\left.\mu_{n}\right\Vert \mu_{m}\right)$
is proved in a similar fasion.
\end{IEEEproof}
In the case where $k=\infty$ Lemma \ref{lem:MinMax} just states
that $\int Y_{m,n}^{max}\,\mathrm{d}\nu=\int Y_{m,n}^{max}\,\mathrm{d}\nu=\infty,$
but with a little modification, we get a much stronger result that
implies the lemma.
\begin{defn}[\cite{Sekhon2021}]
Let $\mu_{1},\mu_{2},\dots,\mu_{n}$ be measures. Then the measure
$\bigvee_{j=1}^{n}\mu_{j}$ is defined by 
\[
\bigvee_{j=1}^{n}\mu_{j}\left(A\right)=\sup\sum_{j=1}^{n}\mu_{j}\left(A_{j}\right)
\]
where the supremum is taken over all disjoint $A_{1},A_{2},\dots,A_{n}$
such that $\bigcup_{j=1}^{n}A_{j}=A.$ Similarly, the measure $\bigwedge_{j=1}^{n}\mu_{j}$
is defined by 
\[
\bigwedge_{j=1}^{n}\mu_{j}\left(A\right)=\inf\sum_{j=1}^{n}\mu_{j}\left(A_{j}\right)
\]
where the infimum is tagen over all disjoint $A_{1},A_{2},\dots,A_{n}$
such that $\bigcup_{j=1}^{n}A_{j}=A.$
\end{defn}
\begin{thm}
Let $\mu$ and $\nu$ denote measures such that $D\left(\mu\Vert\nu\right)<\infty$.
Let $F_{1}\subseteq F_{2}\subseteq\dots\subseteq F_{n}$. Let $Y_{j}=\frac{\mathrm{d}\mu\left(\cdot\mid F_{j}\right)}{\mathrm{d}\nu\left(\cdot\mid F_{j}\right)}$
and let $\mu_{j}$ be the measure given by $\frac{\mathrm{d}\mu_{nj}}{\mathrm{d}\nu}=Y_{j}.$
Let $Y_{m,n}^{min}=\min_{m\leq j\leq n}Y_{j}$ and $Y_{m,n}^{max}=\max_{m\leq j\leq n}Y_{j}.$
Then 
\begin{align*}
D\left(\mu_{n\mid F_{m}}\left\Vert \left(\bigvee_{j=m}^{n}\mu_{j}\right)_{F_{m}}\right.\right) & \leq D\left(\left.\mu_{n}\right\Vert \mu_{m}\right),\\
D\left(\mu_{n\mid F_{m}}\left\Vert \left(\bigwedge_{j=m}^{n}\mu_{j}\right)_{\mid F_{m}}\right.\right) & \leq D\left(\left.\mu_{n}\right\Vert \mu_{m}\right).
\end{align*}
\end{thm}
\begin{IEEEproof}
Let $A_{1},A_{2},\dots,A_{p}$ denote the atoms of $F_{m}.$ Then,
\begin{multline}
D\left(\mu_{m}\left(\cdot\cap A_{i}\right)\left\Vert \left(\bigvee_{j=m}^{n}\mu_{j}\right)\left(\cdot\cap A_{i}\right)\right.\right)\\
\leq D\left(\left.\mu_{n}\left(\cdot\cap A_{i}\right)\right\Vert \mu_{m}\left(\cdot\cap A_{i}\right)\right).
\end{multline}
Now we use that 
\begin{multline}
D\left(\mu_{m\mid F_{m}}\left\Vert \left(\bigvee_{j=m}^{n}\mu_{j}\right)_{F_{m}}\right.\right)\\=\sum_{i=1}^{p}D\left(\mu\left(A_{i}\right)\left\Vert \left(\bigvee_{j=m}^{n}\mu_{j}\right)\left(A_{i}\right)\right.\right)
\end{multline}
and
\[
D\left(\left.\mu_{n}\right\Vert \mu_{m}\right)=\sum_{i=1}^{p}D\left(\left.\mu_{n}\left(\cdot\cap A_{i}\right)\right\Vert \mu_{m}\left(\cdot\cap A_{i}\right)\right)
\]
\end{IEEEproof}
The mixture of the two inequalities is
\begin{multline}
\frac{1}{2}D\left(k\left\Vert \int Y^{max}\,\mathrm{d}\nu\right.\right)+\frac{1}{2}D\left(k\left\Vert \int Y^{min}\,\mathrm{d}\nu\right.\right)\\
\leq D\left(\mu_{n}\Vert\nu\right)-D\left(k\Vert\ell\right)
\end{multline}
and we also get 
\[
m_{\gamma}^{2}\left(\int Y^{max}\,\mathrm{d}\nu,\int Y^{min}\,\mathrm{d}\nu\right)\leq D\left(\mu_{n}\Vert\nu\right)-D\left(k\Vert\ell\right).
\]

\begin{cor}[Radon-Nikodym Theorem]
Assume that $\mu$ and $\nu$ is s-finite and that $D\left(\mu\Vert\nu\right)<\infty$.
If $F_{n}$ is a countable net of finite partitions such that 
$D\left(\mu_{\mid F_{n}}\Vert\nu_{\mid F_{n}}\right)\to D\left(\mu\Vert\nu\right)$
for $n\to\infty$, then the sequence of Radon-Nikodym derivatives
$\frac{\mathrm{d}\mu_{\mid F_{j}}}{\mathrm{d}\nu_{\mid F_{j}}}$converges
pointwise almost surely to a Radon-Nikodym derivative $\rho$ of $\mu$
with respect to $\nu.$
\end{cor}
\begin{IEEEproof}
According to Proposition \ref{prop:Decomp} there exists a measurable
set $B$ such that $\mu\left(\cdot\cap B\right)$ and $\nu\left(\cdot\cap B\right)$
are $\sigma$-finite and such that $\mu\left(\cdot\cap\complement B\right)=\nu\left(\cdot\cap\complement B\right).$
On $\complement B$ the Radon-Nikodym simply equals 1. The set $B$
is $\sigma$-finite so there exists a disjoint decomposition $B=\bigcup_{j=1}^{\infty}A_{j}$
where $\nu\left(A_{j}\right)<\infty.$ Since $D\left(\mu\Vert\nu\right)<\infty$
we also have that $\mu\left(A_{j}\right)<\infty.$ 

Note that $Y_{m,n}^{max}$ is increasing in $n$ and denote the limit
for $n\to\infty$ by $Y_{m,\infty}^{max}.$ Similar $Y_{m,n}^{min}$
is decreasing and the limit will be denoted by $Y_{m,\infty}^{min}.$
Now $Y_{m,\infty}^{max}$is decreasing in $m$ and converges pointwise
to some function $Y_{\infty}^{max}$ and $Y_{m,\infty}^{min}$ is
increasing in $m$ and converges pointwise to some function $Y_{\infty}^{min}.$
Since 
\begin{align*}
D\left(k\left\Vert \int Y_{m,\infty}^{max}\,\mathrm{d}\nu\right.\right) & \leq D\left(\left.\mu_{n}\right\Vert \mu\right),\\
D\left(k\left\Vert \int Y_{m,\infty}^{min}\,\mathrm{d}\nu\right.\right) & \leq D\left(\left.\mu_{n}\right\Vert \mu\right),
\end{align*}
we have that 
\begin{align*}
\int Y_{m,\infty}^{max}\,\mathrm{d}\nu & \to k,\\
\int Y_{m,\infty}^{min}\,\mathrm{d}\nu & \to k,
\end{align*}
for $m\to\infty.$ Hence, $\int\left(Y_{m,\infty}^{max}-Y_{m,\infty}^{max}\right)\,d\nu\to0$,
implying that 
\[
\int Y_{\infty}^{max}\,\mathrm{d}\nu=\int Y_{\infty}^{min}\,\mathrm{d}\nu
\]
so that $Y_{\infty}^{max}=Y_{\infty}^{min}$ $\nu$-almost surely.
Since $Y_{m,\infty}^{min}\leq Y_{n}\leq Y_{m,\infty}^{max}$ we also
have that $Y_{n}\to Y_{\infty}^{max}$ almost surely. 

If $A\in F_{k}$ then $\mu\left(A\right)=\int_{A}Y_{n}\,\mathrm{d}\nu$ for
all $n\geq k$ implying that $\mu\left(A\right)=\int_{A}Y_{\infty}^{max}\,\mathrm{d}\nu$
by the dominated convergence theorem. 

If $A\notin F_{k}$ for any $k$ we may for each $k$ replace each
of the $\sigma$-algebra $F_{k}$ by the $\sigma$-algebra generated
by $F_{k}$ and $A.$ This leads to a sequence of derivates $Z_{k}$
with the same limit almost surely implying that $\mu\left(A\right)=\int_{A}Y_{\infty}^{max}\,\mathrm{d}\nu.$
\end{IEEEproof}

\subsection{Finite divergence as a necessary condition}

Here we will demonstrate that if $D\left(\mu\Vert\nu\right)=\infty$
then there exists a sequence of finite subalgebras and a corresponding
sequence of densities such that the integral of the maximum tends
to $\infty$ for $n\to\infty.$ To simplify the exposition, we will
make some extra assumptions that are not crucial for the result. The
first to prove theorems of this kind were Stein \cite{Stein1969}
and Gundy \cite{Gundy1969}. See \cite{Kuehn2023} for a more recent
presentation.

Assume that $\mu$ and $\nu$ are finite non-atomic measures on a
Polish space and that $\ensuremath{\mu}$ is absolutely continuous
with respect to $\ensuremath{\nu.}$ Let $\rho$ denote the Radon
Nikodym derivative of $\mu$ with respect to $\nu.$ Without loss
of generality, we may assume that $\nu$ is a probability measure.
Further, we may assume that $\nu$ is the Lebesgue measure on the interval
$\left[0,1\right]$. Further, we may assume that $\rho$ is a decreasing
function. 
\begin{lem}
There exists an $\epsilon>0$ such that $\rho\left(x\right)\leq\nicefrac{1}{x}$
for all $x\leq\epsilon.$ 
\end{lem}
\begin{IEEEproof}
Assume that there exists a decreasing sequence $x_{n}$ such that
$\rho\left(x\right)>\nicefrac{1}{x_{n}}$ such that $x_{n}\to0$ for
$n\to\infty.$ Without loss of generality we may assume that $x_{n+1}\leq x_{n}/2.$
We have 
\begin{align*}
\int_{0}^{1}\rho\left(x\right)\,dx & \geq\sum_{n=1}^{\infty}\int_{x_{n+1}}^{x_{n}}\rho\left(x\right)\,\mathrm{d}x\\
 & \geq\sum_{n=1}^{\infty}\int_{x_{n+1}}^{x_{n}}\frac{1}{x_{n}}\,\mathrm{d}x\\
 & =\sum_{n=1}^{\infty}\frac{x_{n}-x_{n+1}}{x_{n}}\\
 & \geq\sum_{n=1}^{\infty}\frac{1}{2}\\
 & =\infty.
\end{align*}
Since we have assumed that $\rho$ was the density of a finite measure
we have obtained a contradiction.
\end{IEEEproof}
\begin{thm}
If $\mu$ and $\nu$ are finite non-atomic Borel measures on a Polish space and
$D\left(\mu\Vert\nu\right)=\infty$ then there exists a sequence of
finite subalgebras such that $\int\max_{i}\rho_{i}\,d\nu=\infty.$
\end{thm}
\begin{IEEEproof}
First we construct a continuous system of subalgebras. Let $\mathcal{F}_{t}$
be the subalgebra generated by the set $\left[0,t\right]$ and by
the $\sigma$-algebra on $\left[t,1\right].$ Then the density of $\mu_{\mid\mathcal{F}_{t}}$
with respect to $\nu_{\mid\mathcal{F}_{t}}$ is equal to 
\[
\rho_{t}\left(x\right)=\begin{cases}
\frac{\int_{0}^{t}\rho\left(x\right)\,\mathrm{d}x}{t}, & x\in\left(0,t\right),\\
\rho\left(x\right), & x\in\left[t,1\right].
\end{cases}
\]
The maximum of these functions is 
\[
\rho_{\max}\left(x\right)=\frac{\int_{0}^{x}\rho\left(s\right)\,\mathrm{d}s}{x}.
\]
Now we have 
\begin{align*}
\int_{0}^{1}\rho_{\max}\left(x\right)\,\mathrm{d}x & =\int_{0}^{1}\frac{\int_{0}^{x}\rho\left(s\right)\,\mathrm{d}s}{x}\,\mathrm{d}x\\
 & =\int_{0}^{1}\left(\int_{0}^{x}\frac{\rho\left(s\right)}{x}\,\mathrm{d}s\right)\,\mathrm{d}x\\
 & =\int_{0}^{1}\left(\int_{s}^{1}\frac{\rho\left(s\right)}{x}\,\mathrm{d}x\right)\,\mathrm{d}s\\
 & =\int_{0}^{1}\rho\left(s\right)\left(\int_{s}^{1}\frac{1}{x}\,\mathrm{d}x\right)\,\mathrm{d}s\\
 & =\int_{0}^{1}\rho\left(s\right)\ln\left(\frac{1}{s}\right)\,\mathrm{d}s.
\end{align*}

Assume that $\rho\left(x\right)\leq\nicefrac{1}{x}$ for $x\leq\epsilon.$
Then 
\begin{multline*}
\int_{0}^{1}\rho_{\max}\left(x\right)\,\mathrm{d}x \\
=\int_{0}^{\epsilon}\rho\left(s\right)\ln\left(\frac{1}{s}\right)\,\mathrm{d}s+\int_{\epsilon}^{1}\rho\left(s\right)\ln\left(\frac{1}{s}\right)\,\mathrm{d}s\\
  \geq\int_{0}^{\epsilon}\rho\left(s\right)\ln\left(\rho\left(x\right)\right)\,\mathrm{d}s+\rho\left(\epsilon\right)\ln\left(\frac{1}{\epsilon}\right)\left(1-\epsilon\right).
\end{multline*}
We see that if \begin{equation}
\int_{0}^{\epsilon}\rho\left(s\right)\ln\left(\rho\left(x\right)\right)\,\mathrm{d}s=\infty
\end{equation}
then $\int_{0}^{1}\rho_{\max}\left(x\right)\,\mathrm{d}x=\infty.$ 

Since $\rho_{\max}\left(x\right)=\sup_{t}\rho_{t}\left(x\right)$
there exists a sequence $t_{n}$ such that $\int_{0}^{1}\sup_{n=1,2,\dots}\rho_{t_{n}}\left(x\right)\,\mathrm{d}x=\infty.$
Therefore $\int_{0}^{1}\sup_{n=1,2,\dots}\rho_{t_{n}}\left(x\right)\,\mathrm{d}x=\infty.$
Hence 
\[
\int_{0}^{1}\sup_{n=1,2,\dots,N}\rho_{t_{n}}\left(x\right)\,\mathrm{d}x\to\infty
\]
for $N\to\infty.$ Now we remark that $\int_{0}^{1}\sup_{n=1,2,\dots,N}\rho_{t_{n}}\left(x\right)\,\mathrm{d}x$
is equal to the integral of the maximum over a finite set of finite
subalgebras.
\end{IEEEproof}

\section{Discussion}

Many of the results in this note have been formulated for valuations
on topological spaces. The results can be extended to a lattice of
locals as used in point-free topology and in computer science. The
basic ideas carry over to this more general setting, but some of the
steps in the proofs require more background theory that is still unpublished,
or at least foreign to most members of the information theory society.
The basic ideas can also be used in a measure-theoretic setup, but
this will, to some extent, only lead to the reformulation of well-known
results.

In general, the theory of concept lattices and valuations is much
closer to ideas about information processing than the usual measure theory
that is built on top of topology. Therefore, the theory of valuations
on concept lattices, may become the theoretical basis for both information
theory, statistics, and probability theory in the future.

\newpage\bibliographystyle{IEEEtran}
\bgroup\inputencoding{latin9}\bibliography{database1}
\egroup

\end{document}